\documentclass[11pt,a4paper]{article}
\usepackage{jinstpub}
\usepackage[hyphenbreaks]{breakurl}
\graphicspath{{fig/}}

\title{Light yield of an undoped CsI crystal coupled directly to a photomultiplier tube at 77~Kelvin}

\author[a,1]{J. Liu\note{Corresponding author.}}
\author[b,c]{M. Yamashita}
\author[a]{A. K. Soma}
\affiliation[a]{Department of Physics, University of South Dakota,\\414 East Clark Street, Vermillion, South Dakota 57069, USA}
\affiliation[b]{Kamioka Observatory, Institute for Cosmic Ray Research, University of Tokyo,\\Higashi-Mozumi, Kamioka, Hida, Gifu, 506-1205, Japan}
\affiliation[c]{Kavli Institute for the Physics and Mathematics of the Universe (WPI), University of Tokyo,\\Kashiwa, Chiba, 277-8582, Japan}
\emailAdd{jing.liu@usd.edu}

\abstract{A light yield of $20.4 \pm 0.8$~photoelectrons/keV was achieved with an undoped CsI crystal coupled directly to a photomultiplier tube at 77~Kelvin. This is by far the largest yield in the world achieved with CsI crystals. An energy threshold that is several times lower than the current dark matter experiments utilizing CsI(Tl) crystals may be achievable using this technique. Together with novel CsI crystal purification methods, the technique may be used to improve the sensitivities of dark matter and coherent elastic neutrino-nucleus scattering experiments. Also measured were the scintillation light decay constants of the undoped CsI crystal at both room temperature and 77~Kelvin. The results are consistent with those in the literature.}


\arxivnumber{1608.06278}
\begin{document}
\maketitle

\section{Introduction}
The thallium doped sodium iodide and cesium iodide at room temperature, NaI/CsI(Tl) in short hereafter, are widely used in direct dark matter search experiments~\cite{dmice16, dama15, kims14}, due to their high scintillation light yields ($\sim$\,50~photons/keV) and relatively low costs. The DAMA experiment observed an annual modulation signal in the 2--6~keV region in their NaI(Tl) scintillators~\cite{dama15}. If it is interpreted with the standard dark matter theory, the observation conflicts with results from experiments using different target materials~\cite{lux16, cdms16, kims14, cogent11}. Many experiments have been proposed to verify the DAMA result using the same material in different environments, including ANAIS~\cite{anais16}, DM-Ice~\cite{dmice16}, KIMS-NaI~\cite{kims16}, COSINE~\cite{cosine16} and SABRE~\cite{sabre15} etc. One of the main difficulties is to reduce the radioactive contamination, $^{40}$K for instance, in NaI powder. The $\sim$3~keV X-rays/Auger electrons from $^{40}$Ar, the daughter of $^{40}$K electron-capture decay, are potential background for dark matter search. A higher light yield of NaI/CsI crystals will lead to a lower energy threshold and better energy resolution. The analysis window can be moved from 2--6~keV to a lower energy region, 1--2~keV for example, where more dark matter signal is expected and the contamination from the $\sim$3~keV peak is negligible.

Sodium doped CsI crystals are used in the CosI experiment~\cite{cosi15} to detect coherent elastic neutrino-nucleus scattering (CEvNS) events. The CEvNS process is important in the evolution of astronomical objects~\cite{janka07}. It can also be used to probe non-standard neutrino interactions~\cite{bar05}, sterile neutrinos~\cite{formaggio12} and nuclear structures~\cite{patton12, patton13}, as well as to monitor the activity of a reactor~\cite{drukier84}. Many experiments along with CosI have been performed or proposed to detect it, for example, TEXONO~\cite{texono16}, CoGeNT using Ge detectors, RED~\cite{red16} using liquid xenon, and COHERENT~\cite{cosi15, coherent15} using all of them, etc. The major difficulty is that the energies of recoiled nuclei are so small that detectors with extremely low energy thresholds are needed.

A straight forward way to lower the energy threshold of a scintillator is to increase its light yield. It was observed that the light yields of undoped NaI/CsI increase rapidly when temperature goes down, and reach the highest point around liquid nitrogen temperature (77~Kelvin at one atmospheric pressure)~\cite{Bonanomi52, Hahn53a, Hahn53, Sciver56, Beghian58, Sciver58, Sciver60, Hsu66, Fontana68, West70, Fontana70, Emkey76, persyk80, Woody90, Williams90, Nishimura95, Wear96, Amsler02, Moszynski03, Moszynski03a, Moszynski05, Moszynski09, Sibczynski10, Sibczynski12, Sailer12}. A theoretical explanation based on quantum statistics was given by Hsu et al. in 1966~\cite{Hsu66}. The observed photons varied with the purity of crystals and light readout methods. Nevertheless, all measurements gave similar or higher yields than those of NaI/CsI(Tl) at room temperature. The highest ones~\cite{Bonanomi52, Sciver56, Moszynski03, Moszynski05} almost reach the theoretical limit deduced from the band gap energy.

Given such high yields, however, the application of undoped NaI/CsI at 77~Kelvin is rare, because the resistivity of the photo-cathode of a photomultiplier tube (PMT), becomes too high at 77~Kelvin. Light guides coupled to PMTs at room temperature were used to circumvent this problem. But light losses in guides and between multiple optical interfaces cancel out the high yield.

The avalanche photodiode (APD) has been used to read out light from undoped NaI/CsI at low temperatures~\cite{Moszynski05,Sibczynski12}. Since it is operated at linear amplification mode, the amplification factor is only at the order of 100 to 1000, much smaller than that of a PMT. This limitation makes it impossible to be used to lower the energy threshold for dark matter or CEvNS detection.

Silicon photomultiplier (SiPM) is basically an array of APDs working in Geiger mode. It can be used to detect individual photoelectrons. It is very noisy at room temperature, but quiet at 77~Kelvin. All make it very suitable as a readout sensor for undoped NaI/CsI at 77~Kelvin. However, there are some issues~\cite{sipm} limiting its usage in dark matter or CEvNS detection.  First of all, the after pulse rate rises very fast below 120~Kelvin. Secondly, the quantum efficiency (QE) reaches the highest point around 500~nm to 600~nm, while the wavelengths of undoped NaI/CsI at 77~Kelvin peak at around 313~nm~\cite{Sibczynski12} and 340~nm~\cite{Nishimura95, Woody90, Amsler02}, respectively. Third, there are optical cross talks between individual APDs in the array. Last, the linear energy response is only available if the number of hit APDs is below a certain percentage of the total number of APDs. Nevertheless, it is an attractive option. The issues may be addressed in a separate paper.

It was tried by Persyk et al. in 1980 to couple an undoped NaI directly to a PMT with metal coated beneath its photo-cathode to keep its resistivity low at 77~Kelvin~\cite{persyk80}. However the PMT (Hamamatsu R878) had very low QE for the 300~nm scintillation light emitted from an undoped NaI at 77~Kelvin.

The performance of PMTs at cryogenic temperatures has been improved gradually over the years. Recently, Hamamatsu Photonics K.K. developed PMT R8778MODAY(AR)~\cite{yamashita10} based on PMT R8778 for the XMASS experiment~\cite{minamino12}. Its QE is approaching 30\% even at 77~Kelvin. Several other PMTs are also available from Hamamatsu to operate at 77~Kelvin, for example, R6041-506 and R11065, etc~\cite{hotta14}. By coupling undoped NaI/CsI crystals directly to those PMTs at 77~Kelvin, a light collecting efficiency higher than previous measurements is expected. An energy threshold lower than those of DAMA and KIMS can be achieved.

An experiment was performed to verify this idea. The light yield of the whole system, including an undoped CsI crystal and a PMT R8778MODAY(AR), was measured at both room temperature and 77~Kelvin and reported here. The data from an undoped NaI crystal is under processing and will be published separately.

\section{Experimental setup}
\label{sec:expt}
\begin{wrapfigure}{r}{0.6\linewidth}
    \includegraphics[width=\linewidth]{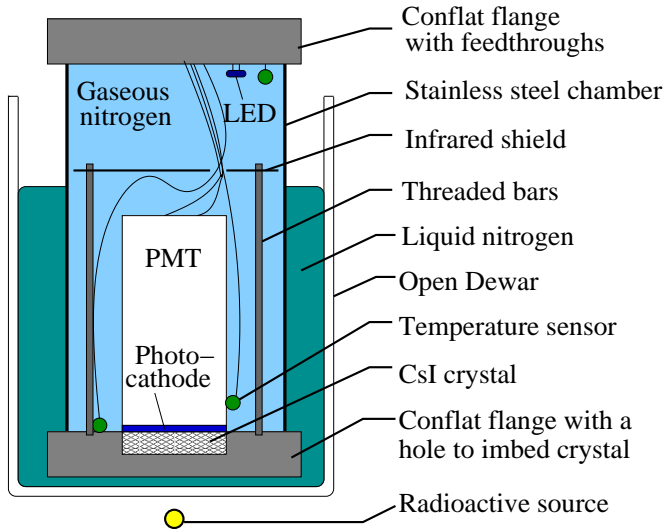}
    \caption{\label{f:setup}Schematic experimental setup (not to scale).}
\end{wrapfigure}

Figure~\ref{f:setup} shows the schematic of the experimental setup (not to scale). Both the PMT and the crystal was housed in a 50~cm long stainless steel chamber, which was vacuumed down to about $10^{-6}$~mbar and then filled with nitrogen gas up to 0.17~MPa (absolute). The chamber was placed inside an open dewar so that it could be cooled down by pouring liquid nitrogen into the dewar. The temperature profile inside the chamber was controlled by varying the liquid nitrogen level and monitored by three PT100 temperature sensors. One sensor was pressed against the bottom Conflat flange of the chamber, another was mounted to the side surface of the PMT close to the photo-cathode, the last one was hanging right below the top Conflat flange, close to two LEDs used to calibrate the PMT light response. The light emission spectra of the LEDs were almost Gaussian and peaked at 420~nm and 300~nm, respectively. All electronic and gas feedthroughs were welded into a 6-inch Conflat flange on the top of the chamber.

\begin{figure}[htpb]
    \includegraphics[width=0.32\linewidth]{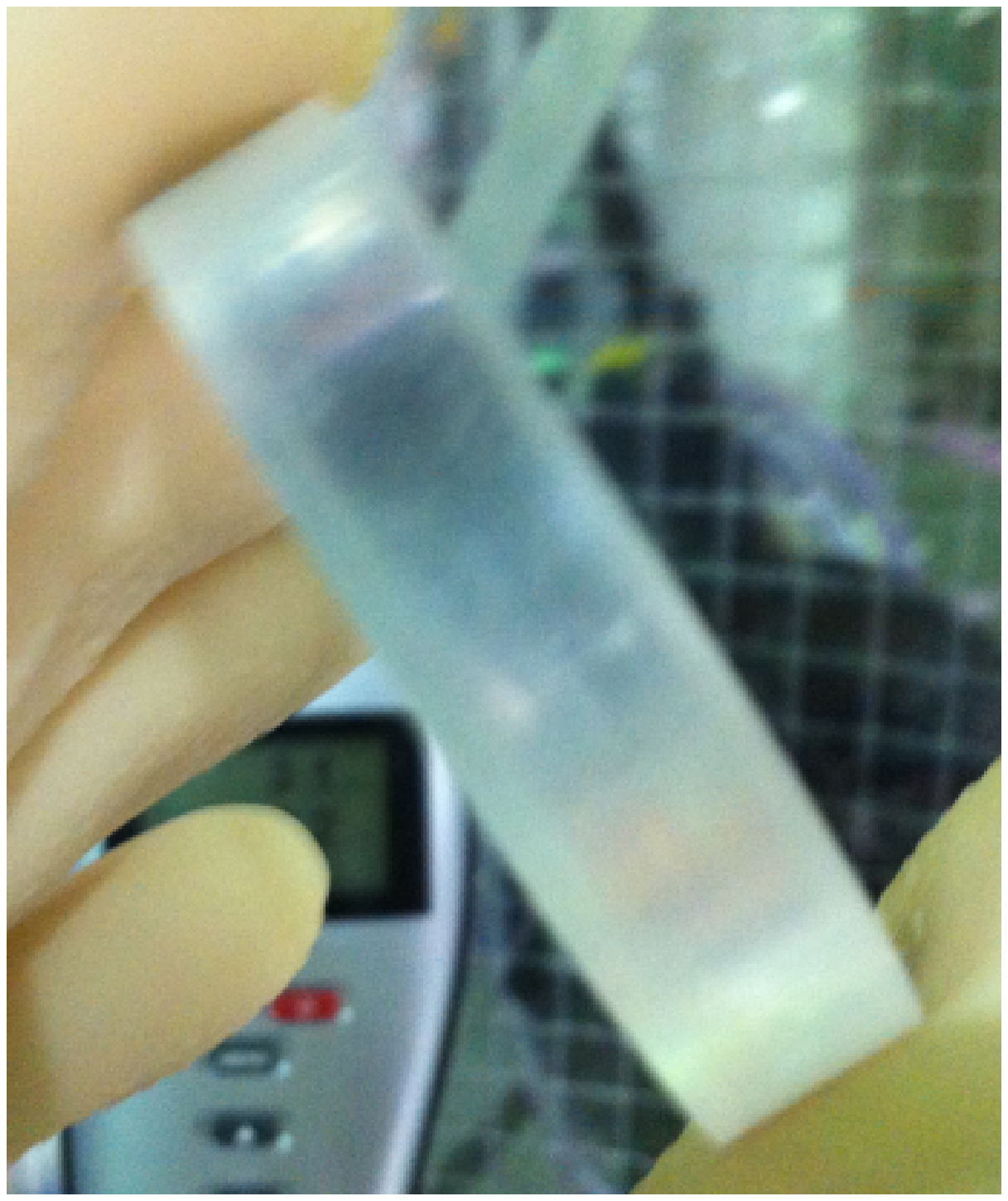}\hfil
    \includegraphics[width=0.32\linewidth]{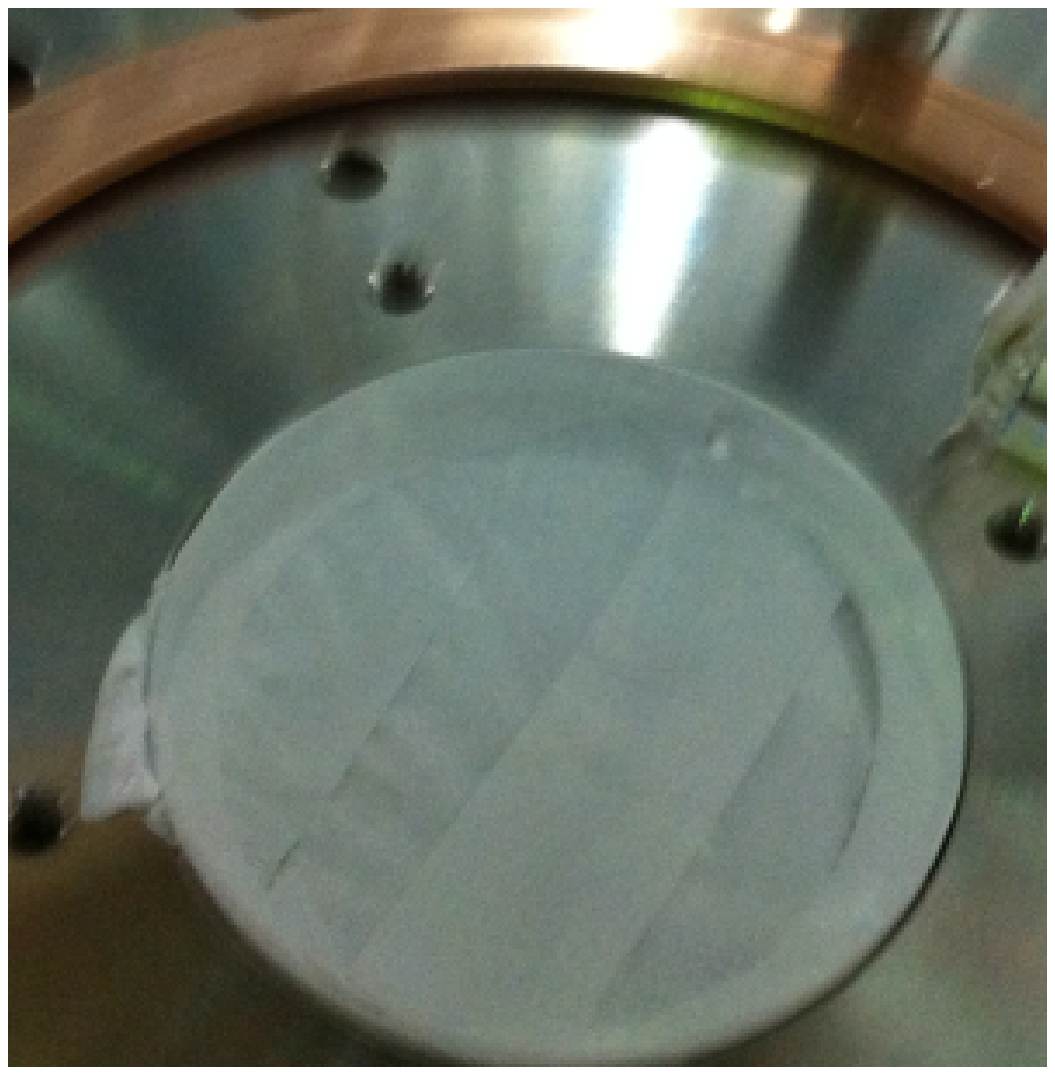}\hfil
    \includegraphics[width=0.32\linewidth]{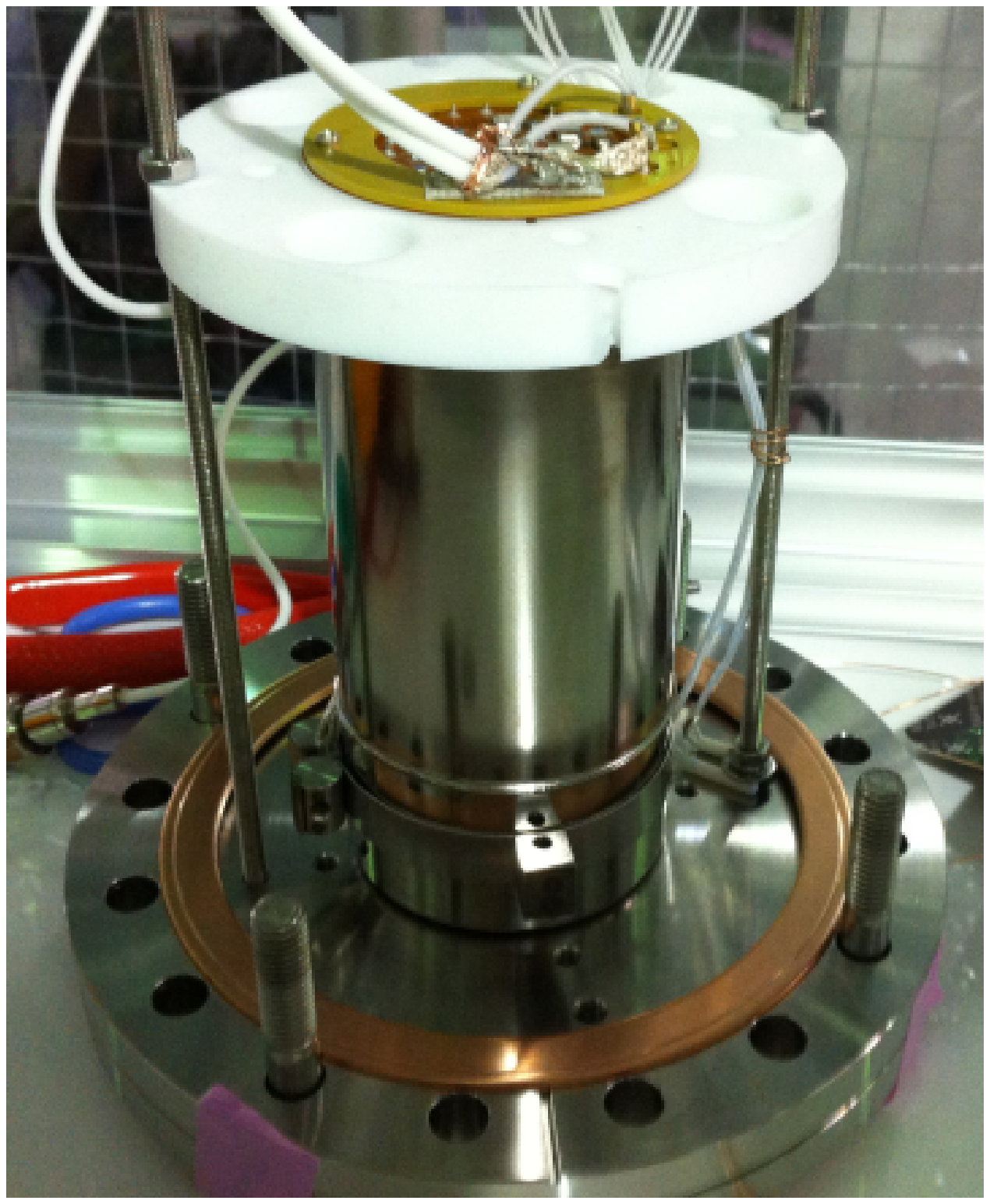}
    \caption{\label{f:csi}Left: undoped cylindrical CsI crystal with its side surfaces roughly polished. Middle: the crystal embedded into the customized Conflat flange with its bottom and side surfaces wrapped with Teflon tape. Right: PMT mounted on top of the crystal with its photo-cathode facing downward.}
\end{figure}

The undoped cylindrical CsI crystal has a height of 1~cm and a radius of 1~inch. Its top and bottom surface were nicely polished. The side surface was roughly polished as shown in the left most picture in figure~\ref{f:csi}. Its side and bottom surfaces were wrapped with Teflon tapes to reflect scintillation light.  The crystal was embedded into a hole at the center of a customized 6-inch Conflat flange at the bottom of the chamber. The dimensions of the hole were slightly larger than those of the crystal so that the Teflon tape wrapped crystal could be easily placed in as shown in the middle picture of figure~\ref{f:csi}. The PMT was mounted upside down with its $\sim$2~mm thick quartz window tightly pushed against the top surface of the crystal to ensure good optical contact without glue, as shown in the right most picture in figure~\ref{f:csi}.  The photo-cathode was on the inner surface of the window. Optical photons from the LEDs could reach the photo-cathode through the side of the quartz window.

Most of the assembly process of the system was done in a glove box flushed with dry nitrogen gas. The relative humidity was below 3\% at 18$^\circ$C during the process. Limited by the size of the glove box, the connection of cables to the top flange was done outside the glove box with a relative humidity around 45\%.

The PMT photo-cathode was grounded, the anode was biased to 1,200~V (recommended by Hamamatsu Photonics K.K.) for the measurements at room temperature. The standard cathode-grounding high-voltage dividing circuit was used. The signal coming out of the circuit was fed to a fan-in/fan-out NIM module. One copy was sent to a standard NIM discriminator to generate the trigger signal, the other was sent to a CAEN V1751 waveform digitizer with a sampling rate of 1~GHz and a bandwidth of 500~MHz. The dynamic range of the digitizer is 1~V, the resolution is 10 bits. One digital count given by the Analog-to-Digital Converter (ADC count) hence represents 1~V/$2^{10} = 0.98$~mV.  The digitizer was triggered externally by either the discriminator or a TTL clock signal, which was also used to trigger the LED light emission.

A slightly lower voltage, 1,020~V, was used for the measurements at 77~Kelvin, since the scintillation signal became so strong that the PMT gain had to be lowered to avoid saturation. The electron collection efficiency was measured to be the same as that at 1,200~V.  A Phillips Scientific NIM Model 776 had to be used to amplify the PMT signal by a factor of 10 before it was sent to the digitizer in the measurement of the PMT's single photoelectron (PE) response at 77~Kelvin, otherwise the single PE distribution could not be clearly separated from the pedestal noise distribution.

\section{Single-photoelectron response of the PMT}
\label{s:1pe}
A 500~Hz TTL clock signal was used to trigger both the waveform digitizer and the LED power supply. An 800~ns long waveform was recorded after each trigger. The intensities of the LEDs were tuned such that only one or zero photon struck the PMT photo-cathode during a data acquisition window most of the time.

\begin{figure}[htpb]
  \includegraphics[width=0.49\linewidth]{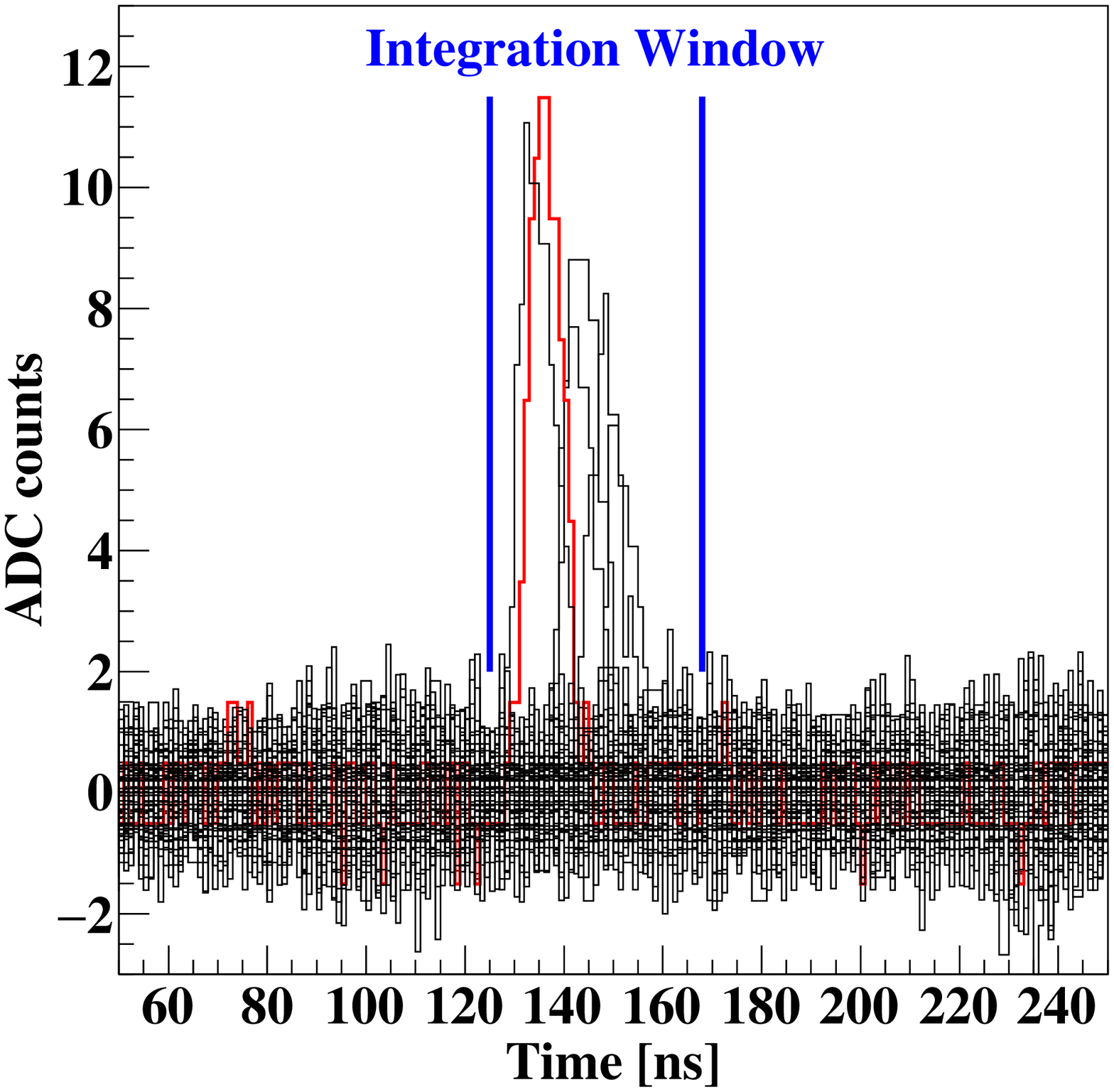}\hfill
  \includegraphics[width=0.49\linewidth]{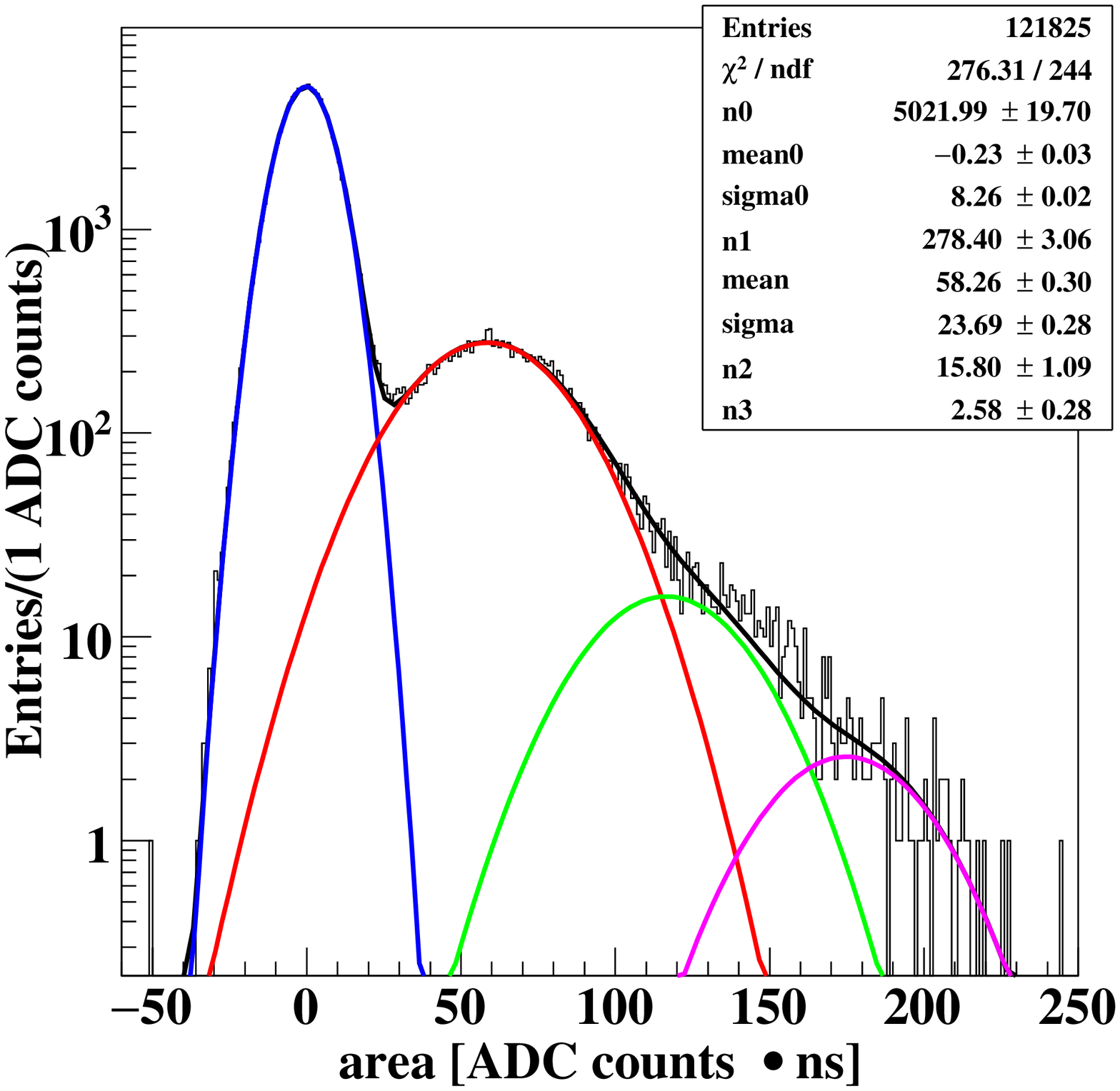}
  \caption{\label{f:c1p3} Left: 50 consecutive waveforms overlapped with each other. Right: single PE response of the PMT at a bias voltage of 1,020~V with an external amplifier (amplification factor: 10).}
\end{figure}

The left plot of figure~\ref{f:c1p3} shows 50 consecutive PMT waveforms acquired at 77~Kelvin, one is colored in red for a better display of a single PE pulse. The pulse in a waveform, if there are any, is distributed within (125, 165)~ns. The integration of each waveform was performed from 125~ns to 165~ns. Its distribution is shown in the right plot of figure~\ref{f:c1p3}.  The blue Gaussian function was used to fit the pedestal distribution around zero. The red Gaussian function was used to fit the single PE peak. The green and magenta Gaussian functions were used to fit the second and the third PE distributions, respectively. The widths of the second and the third PE peaks were fixed to that of the single PE peak, and their respective means were fixed to 2 and 3 times of that of the single PE peak.

The measurement was repeated several times. The results are summarized in Table~\ref{t:1PE}. The means of the single PE distributions, $M_\text{1PE}$ (divided by 10 if the amplifier was used), are listed in the last column of the table.
\begin{table}[htbp]
  \centering
  \caption{\label{t:1PE} Summary of PMT single PE response measurements.}
  \smallskip
  \begin{tabular}{c|c|c|c|c}
    \hline
    Temperature   &Temperature    &PMT high   &Wavelength  &Mean of single PE distri-\\
    of PMT [$^\circ$C]&of LED [$^\circ$C] &voltage [V]&of LED [nm] &bution [ADC counts$\cdot$ns]\\
    \hline
    -191.4 & 8.6 & 1,020 & 470 & 5.849  $\pm$ 0.030\\
    -191.4 & 8.8 & 1,020 & 300 & 5.489  $\pm$ 0.077\\
    -192.7 & 17.5& 1,020 & 300 & 5.068  $\pm$ 0.083\\
    18.3   & 17.9& 1,020 & 300 & 3.835  $\pm$ 0.041\\
    18.1   & 17.8& 1,200 & 300 & 16.637 $\pm$ 0.294\\
    -192.7 & 17.5& 1,200 & 300 & 23.700 $\pm$ 0.471\\
    \hline
  \end{tabular}
\end{table}

Several things in table~\ref{t:1PE} are worthy of notice. First, $M_\text{1PE}$ decreases as the applied high voltage decreases. This is expected from the working principle of a PMT. Second, given the same high voltage, $M_\text{1PE}$ increases as the temperature decreases. This may have several causes, including changes of the size, location and properties of PMT dinodes, as well as the resistances of the resistors in the PMT high voltage distributing circuit over temperature. Third, three measurements at 1,020~V in liquid nitrogen give different $M_\text{1PE}$. The standard deviation is 5.9\%. This is too large to be explained by the statistical uncertainties listed in the table. It can neither be explained by the difference of the LED wavelength, which can only affect the QE instead of the PMT gain. The reason is still under investigation. The mean of the first three measurements with 5.9\% uncertainty, $\bar{M}_\text{1PE} = 5.468 \pm 0.32$, was used to convert (ADC counts $\cdot$ ns) to the average number of PE in the energy calibration measurements at 77~Kelvin using the following equation,
\begin{equation}
  \text{(number of PE)} = \text{(ADC counts} \cdot \text{ns)}/\bar{M}_\text{1PE}.
  \label{e:m1pe}
\end{equation}
$M_\text{1PE}$ shown in the second last line in table~\ref{t:1PE} was used as the conversion factor for the energy calibration measurements at room temperature.

$M_\text{1PE}$ can be measured with another method. The number of PE collected from a photo-cathode obeys the Poisson distribution given a fixed intensity of a light source. According to the central limit theorem, when the expected value $\lambda_\text{PE}$ of the Poisson distribution is large enough (more than 100 PE for example), the distribution can be approximated by a Gaussian distribution with its mean $\mu_\text{PE} = \lambda_\text{PE}$ and its variance $\sigma_\text{PE} = \sqrt\lambda_\text{PE}$. Obviously,
\begin{equation}
  \mu_\text{PE} = \lambda_\text{PE} = \sigma^2_\text{PE}.
  \label{e:mspe}
\end{equation}
Similar to eq.~\ref{e:m1pe},
\begin{equation}
  \mu_\text{PE} = \mu_\text{ADC}/M_\text{1PE} \text{ and }
  \sigma_\text{PE} = \sigma_\text{ADC}/M_\text{1PE},
  \label{e:msadc}
\end{equation}
where, $\mu_\text{ADC}$ and $\sigma_\text{ADC}$ are the mean and variance of Gaussian distribution in units of (ADC counts $\cdot$ ns) instead of number of PE. $M_\text{1PE}$ can be calculated combining eq.~\ref{e:mspe} and eq.~\ref{e:msadc}:
\begin{equation}
  (\mu_\text{ADC}/M_\text{1PE}) = (\sigma_\text{ADC}/M_\text{1PE})^2
  \Longrightarrow M_\text{1PE} = \sigma^2_\text{ADC}/\mu_\text{ADC}.
  \label{e:mnpe}
\end{equation}

\begin{figure}[htpb]
  \includegraphics[width=0.49\linewidth]{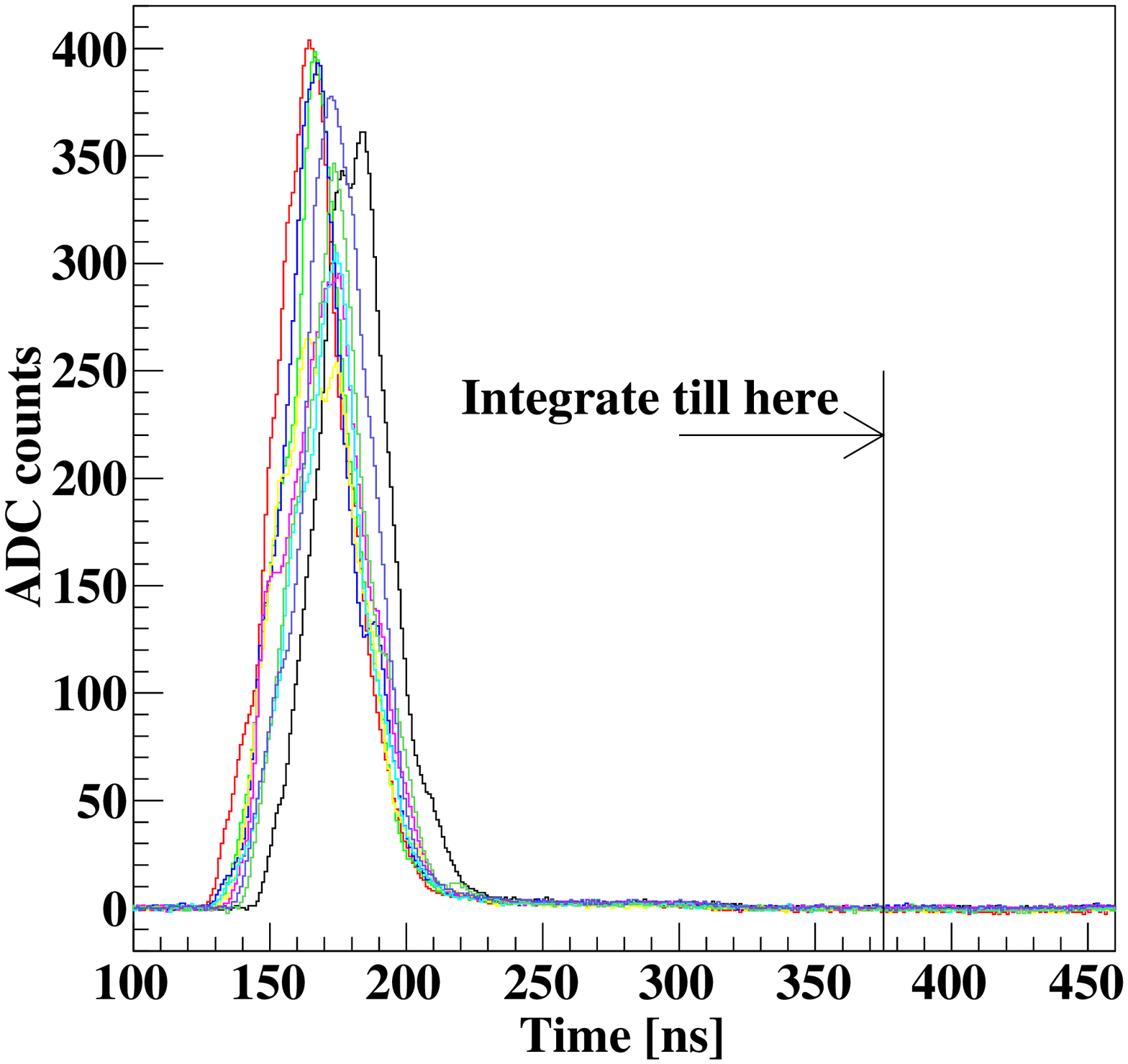}\hfill
  \includegraphics[width=0.49\linewidth]{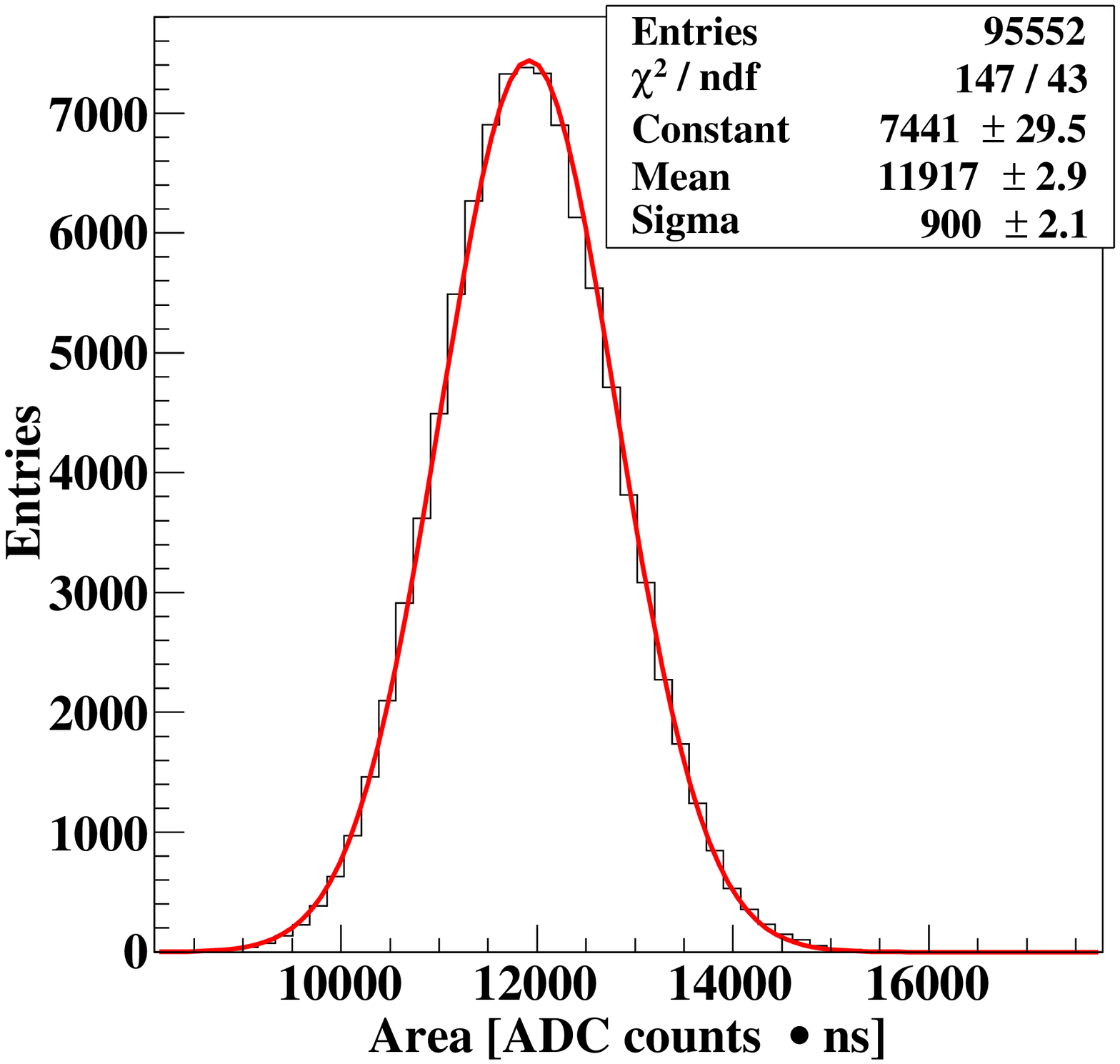}
  \caption{\label{f:hpe}Left: Several consecutive waveforms taken with high LED intensity overlapped with each other. Right: Distribution of the waveforms integration fitted with a Gaussian function.}
\end{figure}

This method was used to double check $M_\text{1PE}$ measured from the single PE distribution shown in figure~\ref{f:c1p3}. The intensity of the light emitted from the LED was raised to a value that is high enough so that the minimal number of PE collected was more than 100. Several waveforms and the (ADC counts $\cdot$ ns) spectrum taken with this setup are shown in figure~\ref{f:hpe}. The spectrum was fitted with a Gaussian function, the mean and variance of which were used to calculate $M_\text{1PE}$ using eq.~\ref{e:mnpe}. $M_\text{1PE}$ obtained using this method is about 15\% larger than that obtained from the single PE measurement. This might be partially due to the fact that the Gaussian distribution is only an approximation of the Poisson distribution, which is reflected from the large $\chi^2$ of the fitting as well. Nevertheless, it indicates that $M_\text{1PE}$ from the single PE measurement would not deviate significantly from the true value.

\section{Energy calibration of the system}
\begin{figure}[htpb]
 \includegraphics[width=\linewidth]{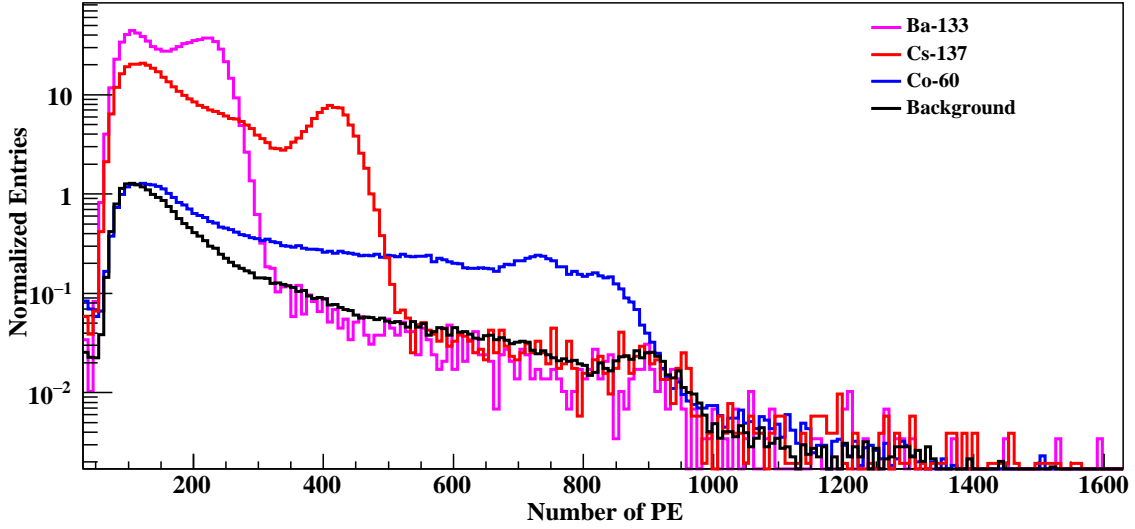}
  \caption{\label{f:srcrt} Energy spectra in terms of number PE at room temperature.}
\end{figure}

The energy spectra taken at room temperature with and without radioactive sources are shown in figure~\ref{f:srcrt}.  The trigger rate for background, $^{137}$Cs, $^{133}$Ba and $^{60}$Co data taking was 187~Hz, 280~Hz, 300~Hz and 330~Hz, respectively. The $x$-axis has been converted from (ADC counts $\cdot$ ns) to number of PE using the method described in section~\ref{s:1pe}. The spectra are normalized using the event rate above 1000 PE. The peak around 900~PE in the background spectrum corresponds to the 1.46~MeV $\gamma$-ray radiation from $^{40}$K in the crystal. The 1.17~MeV and 1.33~MeV peaks of $^{60}$Co can not be resolved due to the poor resolution and appears around 750~PE. The 662~keV peak of $^{137}$Cs and 356~keV peak of $^{133}$Ba can be seen around 450~PE and 250~PE, respectively.

\begin{figure}[htpb]
  \includegraphics[width=\linewidth]{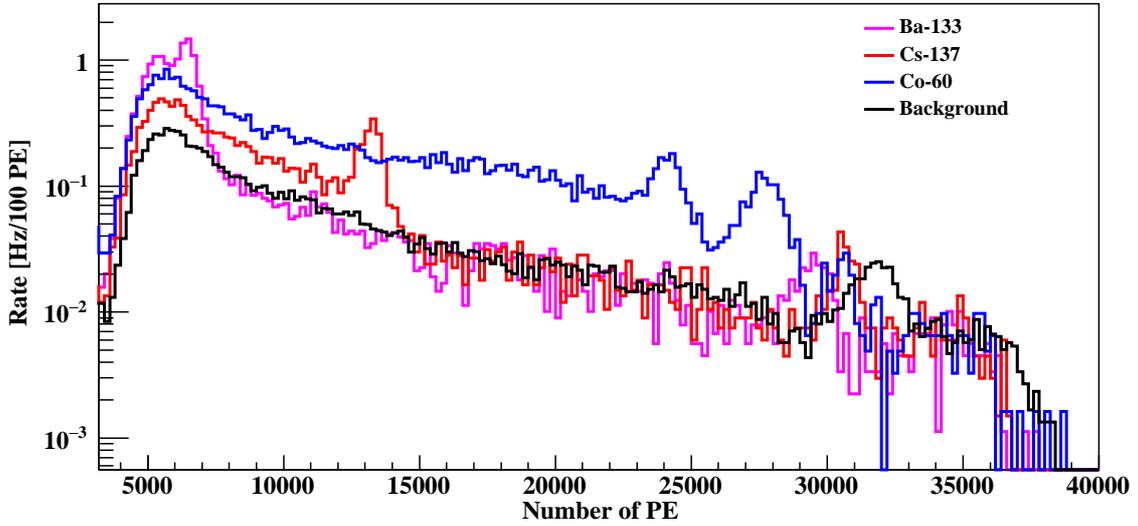}
  \caption{\label{f:srclnt} Energy spectra in terms of number of PE at 77~Kelvin.}
\end{figure}

The energy spectra taken at 77~Kelvin with and without radioactive sources are shown in figure~\ref{f:srclnt}. 
The trigger rate for background, $^{137}$Cs, $^{133}$Ba and $^{60}$Co data taking was 8~Hz, 16~Hz, 21~Hz and 33~Hz, respectively. The $x$-axis has been converted from (ADC counts $\cdot$ ns) to number of PE using the method described in section~\ref{s:1pe}. The spectra are normalized using the life times of the measurements. Compared to the spectra at room temperature, all peaks move to much higher number of PE, indicating a dramatic improvement of the light yield. The 1.17~MeV and 1.33~MeV peaks of $^{60}$Co are clearly separated, indicating a significant improvement of the energy resolution. The 1.46~MeV peak from $^{40}$K can now be seen in all spectra. However, they do not overlap with each other perfectly, indicating an about 5\% change of the gain of the PMT during the measurement, which is reflected also in the variation of $M_\text{1PE}$ mentioned in section~\ref{s:1pe}.

\section{Light yield of the system}
\begin{figure}[htpb]
  \includegraphics[width=0.49\linewidth]{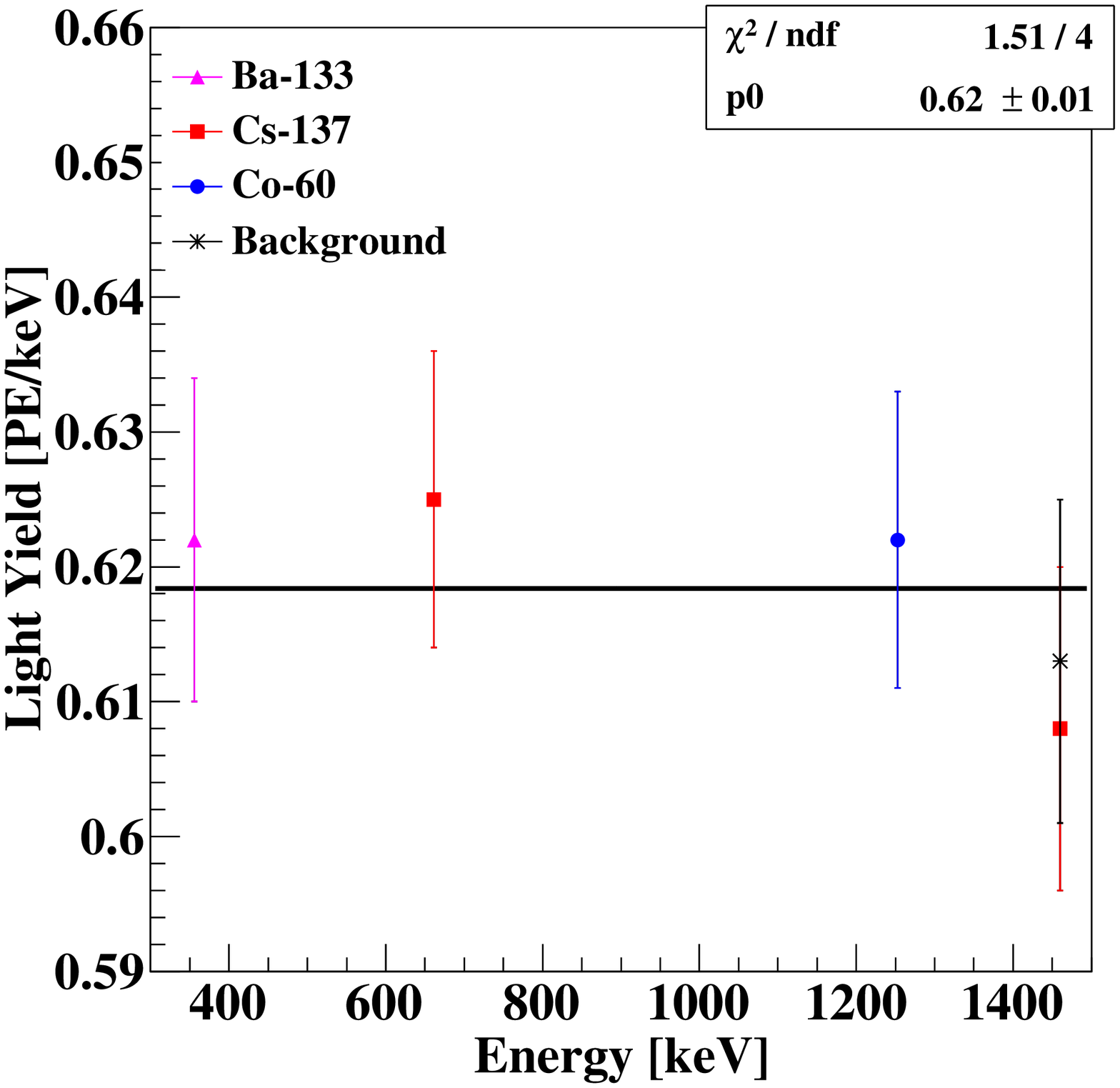}\hfill
  \includegraphics[width=0.49\linewidth]{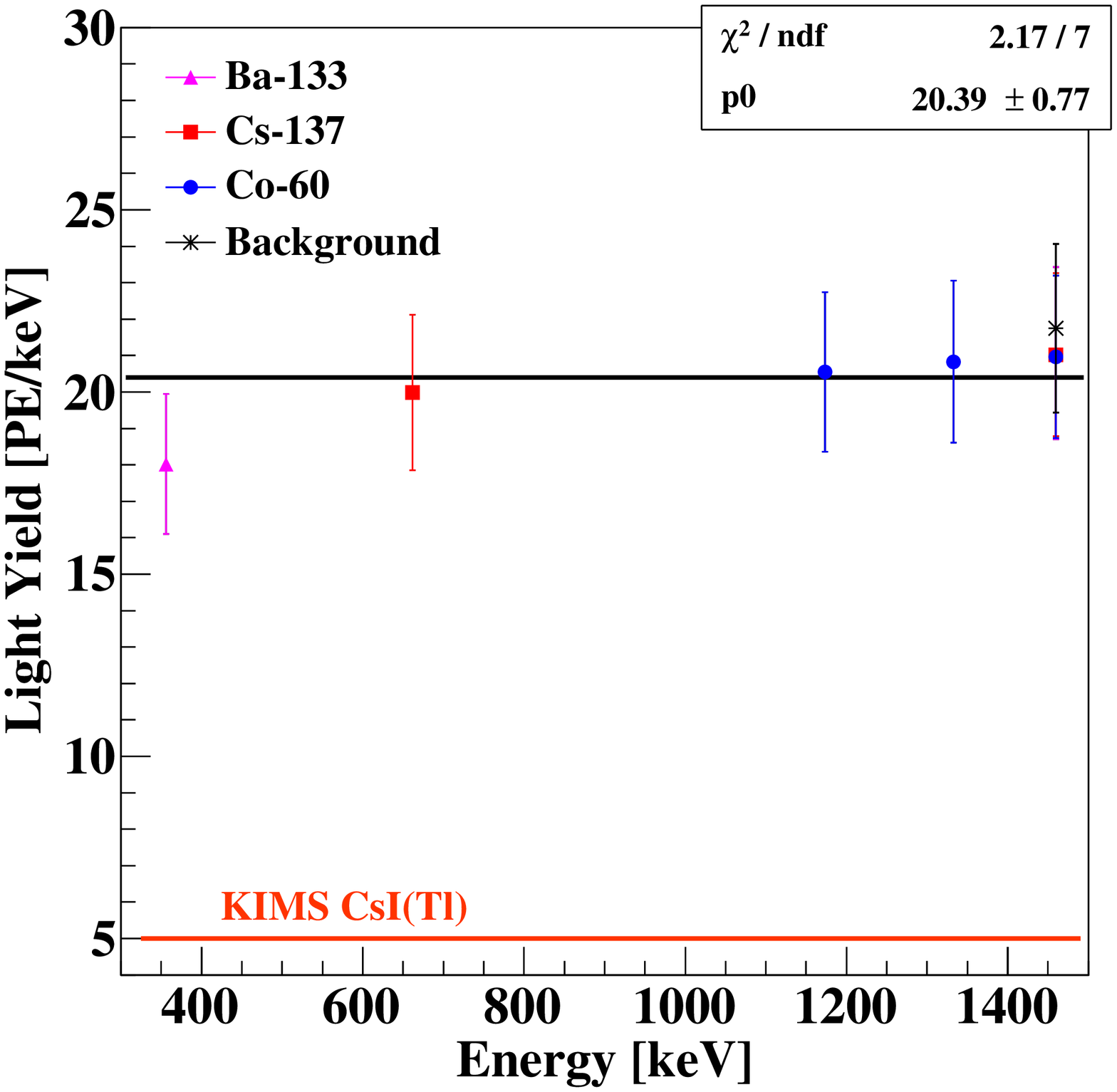}
  \caption{\label{f:ly} Light yield of the entire detection system measured at room temperature (left) and 77~Kelvin (right).}
\end{figure}

Peaks in the energy spectra are summarized in table~\ref{t:rPE}. Their positions are used to calculated the light yield of the detection system using the following equation:
\begin{equation}
  \text{Light yield [PE/keV]} = \text{Peak position [number of PE]}/\text{Peak energy [keV]}.
\end{equation}
The results are shown in figure~\ref{f:ly}. The uncertainty of each data point is mainly due to the uncertainty of $M_\text{1PE}$ used to convert the $x$-axes of the energy spectra from (ADC counts $\cdot$ ns) to number of PE.

\begin{table}[htbp]
  \centering
  \caption{\label{t:rPE} Summary of $\gamma$-ray peaks in the energy spectra.}
  \smallskip
  \begin{minipage}{\linewidth}
  \begin{tabular}{c|c|c|c|c|c}
    \hline
   Radioactive   &Energy   &\multicolumn{2}{c|}{77 K}  &\multicolumn{2}{c}{Room Temperature} \\
   \cline{3 -4}  \cline{4 -6}
      Isotopes   & (MeV)          &Mean (PE) & Variance (PE) & Mean (PE) & Variance (PE)  \\
    \hline
    $^{133}$Ba   & 0.356 & 6415.5   & 359.3 & 221.2  & 48.7\\
    $^{137}$Cs   & 0.662 & 13222.8  & 332.3 & 413.5  & 30.7\\
    $^{60}$Co    & 1.17  & 24110.4  & 483.8 & 729.7  & 61.8\\
                 & 1.32  & 27757.3  & 543.3 &        &      \\
    $^{40}$K~\footnote{in background measurement}& 1.46  &31750.6 & 681.2 & 895.0 & 59.0\\
    $^{40}$K~\footnote{in $^{60}$Co measurement} &       &30604.5 & 307.8 &       &      \\
    $^{40}$K~\footnote{in $^{137}$Cs measurement}&       &30690.7 & 229.0 & 887.6 & 45.2 \\
    $^{40}$K~\footnote{in $^{133}$Ba measurement}&       &30763.7 & 602.9 &       &      \\
    \hline
  \end{tabular}
  \end{minipage}
\end{table}

The average light yield of $0.618\pm0.005$~PE/keV and $20.4\pm0.8$~PE/keV was obtained at room temperature and 77~Kelvin, respectively, by fitting the data points with a straight line, shown as the horizontal black line in each plot in figure~\ref{f:ly}. The small $\chi^2$ of the fittings may indicate some degrees of overestimation of the uncertainty of each data point. The light yield at 77~Kelvin is about 4 times larger than that has been achieved by the KIM-CsI experiment~\cite{kims14}, which is plotted as the red line in the right plot of figure~\ref{f:ly} for comparison. There seems to be a systematic increase of the light yield as the energy increases as shown in the right plot of figure~\ref{f:ly}. This may indicate a slight nonlinearity in the energy response of the undoped CsI crystal at 77~Kelvin. However, limited by the large uncertainty of each data point, no quantitative conclusion can be drawn.

\section{Quantum efficiency of the PMT}
The light yield of the crystal, $\text{LY}_\text{crystal}$, can be calculated from the light yield of the system, $\text{LY}_\text{system}$, shown in the previous section, using the following equation:
\begin{equation}
  \text{LY}_\text{crystal} =  \text{LY}_\text{system} / \epsilon / \text{QE},
  \label{e:ly}
\end{equation}
where, $\epsilon$ is the light collection efficiency, QE is the quantum efficiency of the PMT. The light collection efficiency would not change much if the reflectivity of the Teflon tape does not change much over temperature. To make sure that the increase of $\text{LY}_\text{system}$ is indeed due to the increase of $\text{LY}_\text{crystal}$, the stability of the QE over temperature has to be checked.

\begin{wrapfigure}{r}{0.5\linewidth}
    \includegraphics[width=\linewidth]{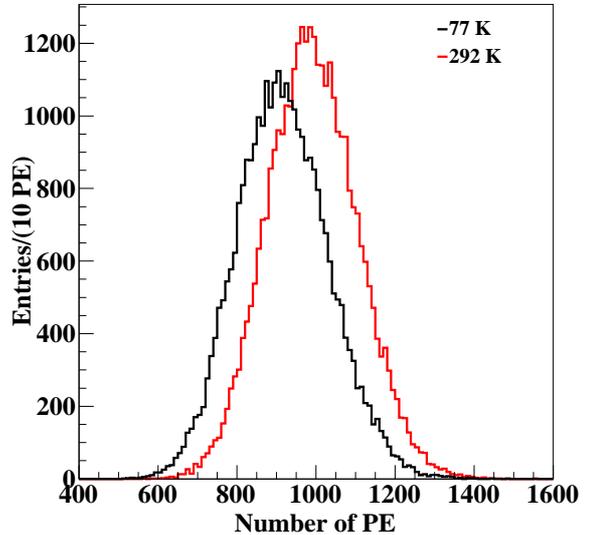}
    \caption{\label{f:qe} Number of PE detected at room temperature (red) and 77~Kelvin (black) with a fixed LED intensity.}
\end{wrapfigure}

This was done by comparing the numbers of PE detected at different temperatures with a fixed LED light intensity. As shown in figure~\ref{f:setup}, the LED ambient temperature was recorded by a PT100 sensor mounted close by. The values are summarized in table~\ref{t:1PE}. The difference is less than 10~$^\circ$C. The LED emission spectrum should not change much under such a small temperature variation according to the specification. The reason why the LED could be kept at room temperature while the PMT was around 77~Kelvin is because the 50~cm long stainless steel chamber thermally decoupled its cool bottom and warm top. This is not obvious as figure~\ref{f:setup} is not to scale.

The result obtained with the 300~nm LED is shown in figure~\ref{f:qe}. The red line shows the number of PE distribution measured at root temperature, the black one at 77~Kelvin. The QE decreases as the temperature decreases. The relative difference between the means of the two distributions in figure~\ref{f:qe} is 7.58\%. The study with the 470~nm LED resulted in a relative difference of 9.24\%.  The QE of the PMT quoted by Hamamatsu Photonics K.K. at room temperature is 29.5\% at 300~nm. The QE at 77~Kelvin is then $29.5\%(1-7.58\%) = 27.3\%$. It can be seen that the increase of $\text{LY}_\text{system}$ is purely due to the increase of $\text{LY}_\text{crystal}$.

\section{Scintillation decay constants of the crystal}
Events in the 1.46~MeV $^{40}$K peaks in the background spectra taken at both room temperature and 77~Kelvin were used to measure the scintillation decay constants of the crystal. The results were compared to those in the literature to make sure that the crystal under study was indeed an undoped CsI.

\begin{figure}[htpb]
  \includegraphics[width=0.49\linewidth]{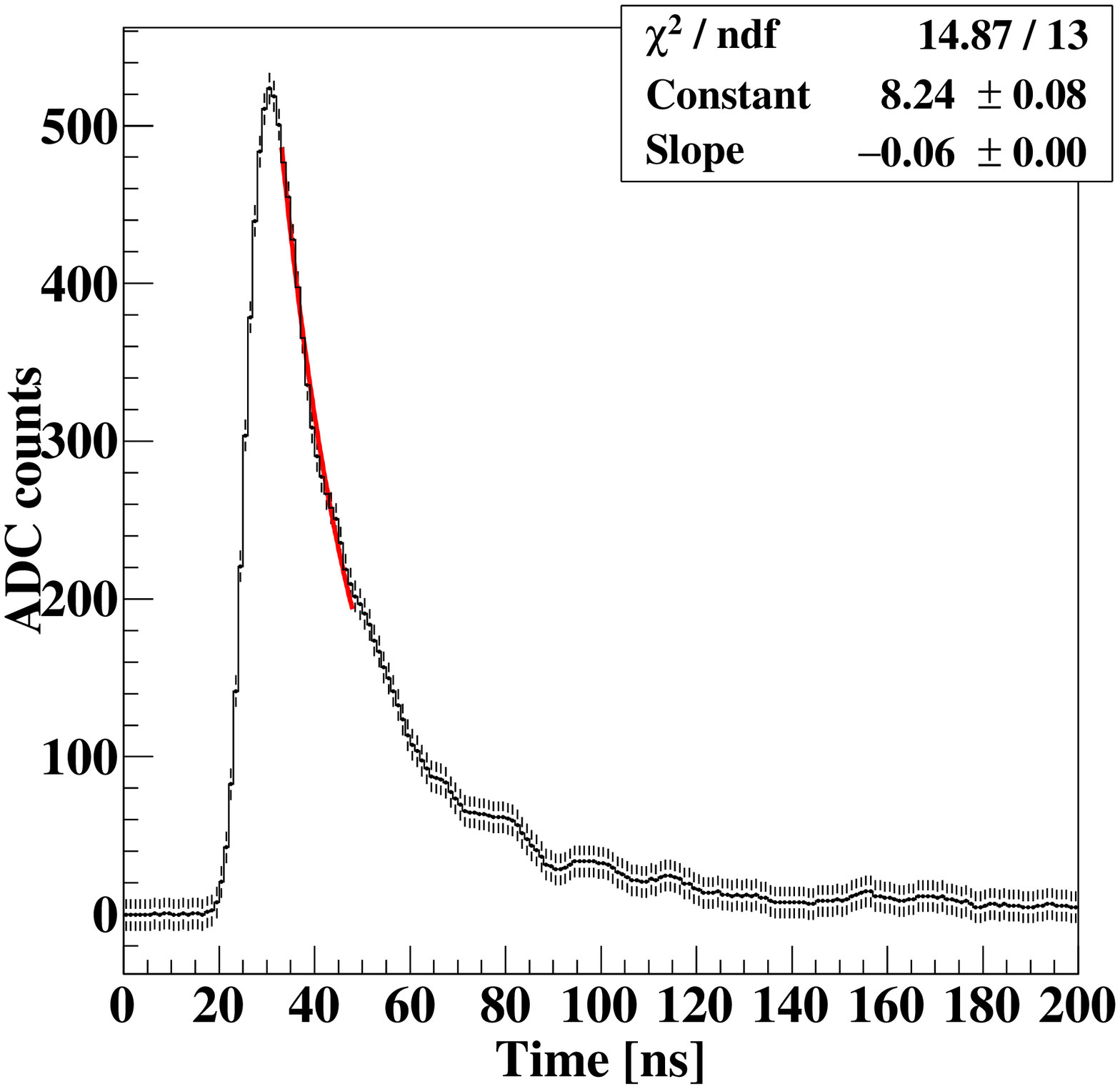}\hfill
  \includegraphics[width=0.49\linewidth]{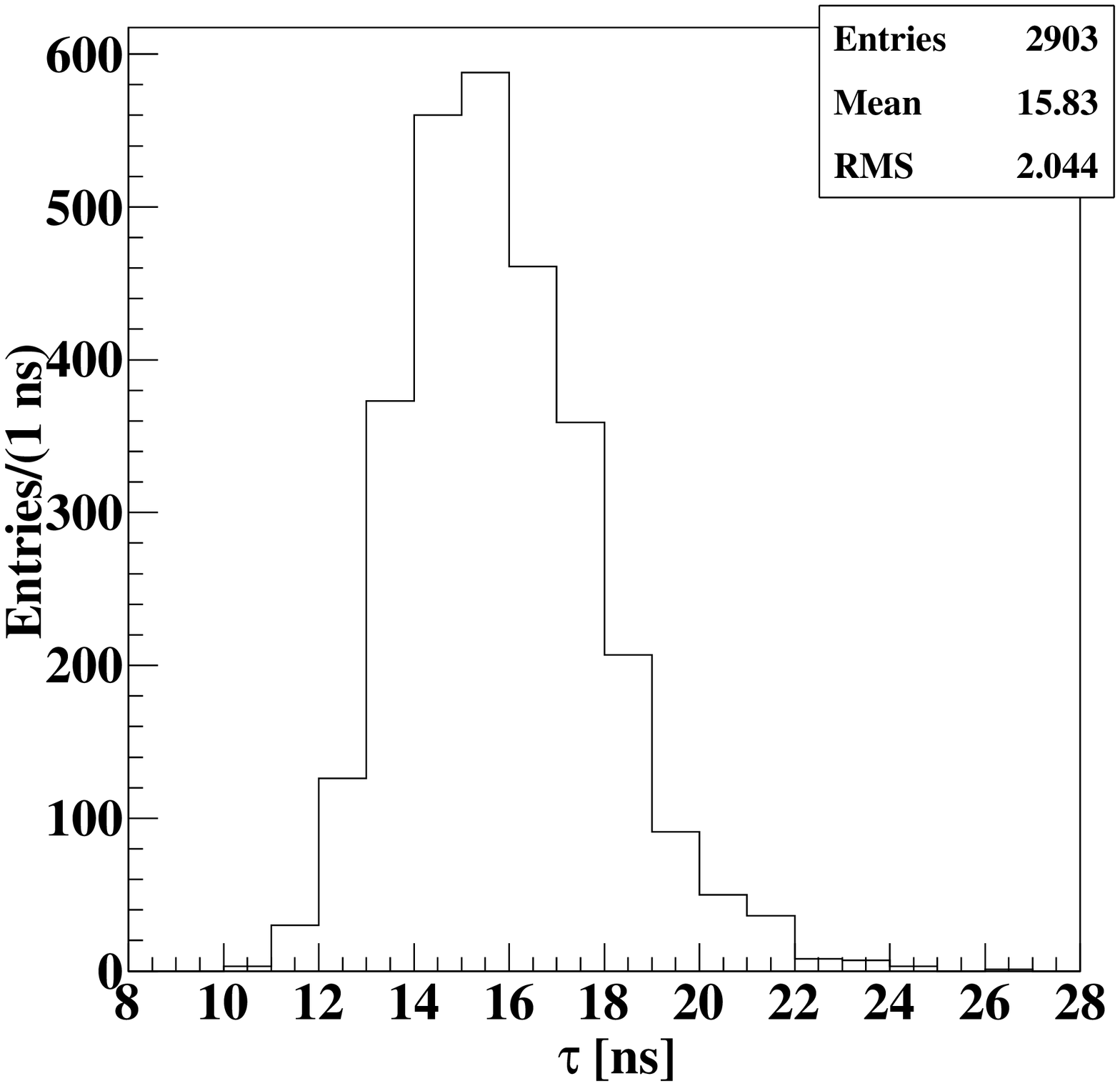}
  \caption{\label{f:rtpls} Left: a typical light pulse taken at room temperature fitted with eq.~\ref{e:exp} to measure the decay constant. Right: distribution of the decay constant defined in eq.~\ref{e:tau}.}
\end{figure}

A typical light pulse taken at room temperature is shown in the left plot of figure~\ref{f:rtpls}. Also shown is an exponential function, 
\begin{equation}
  \text{Constant}\times e^{\text{Slope}\cdot x},
  \label{e:exp}
\end{equation}
fitted to a 15-ns window, 3~ns after the maximum of the pulse. A sinusoidal fluctuation of $\sim$10 ADC counts can be seen in the 45--160~ns time window and was assigned as the uncertainty to each waveform sample. The distribution of the decay constant,
\begin{equation}
  \tau = -1/\text{Slope},
  \label{e:tau}
\end{equation}
obtained from all events is shown in the right plot of figure~\ref{f:rtpls}. The distribution has a mean of 15.83~ns and an RMS of 2.04~ns.

\begin{figure}[htpb]
  \includegraphics[width=\linewidth]{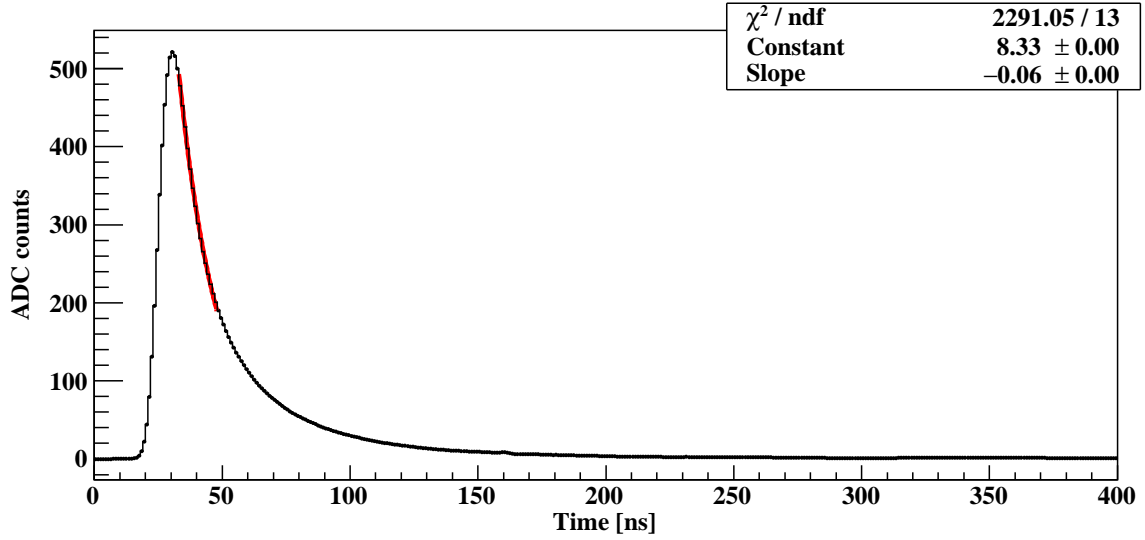}
  \caption{\label{f:rtsum}Averaged waveform of 2903 events at room temperature fitted with eq.~\ref{e:exp}.}
\end{figure}

To average out the electronic noise presented in individual pulses, waveforms from 2903 events were summed up together. The averaged waveform is shown in figure~\ref{f:rtsum}. The standard deviation of the ADC counts of the first 25 samples is 0.26 and was used as the uncertainty of each sample in the waveform. The same exponential fitting as of figure~\ref{f:rtpls} was performed and the decay constant obtained this way was $\tau = 15.63 \pm 0.02$~ns, consistent with the one evaluated from individual pulses. The large $\chi^2$ of the fitting indicates that some systematic uncertainties have not been taken into account. Contamination from other decay components in the fitted time window may be one of them.

\begin{figure}[htpb]
 \includegraphics[width=\linewidth]{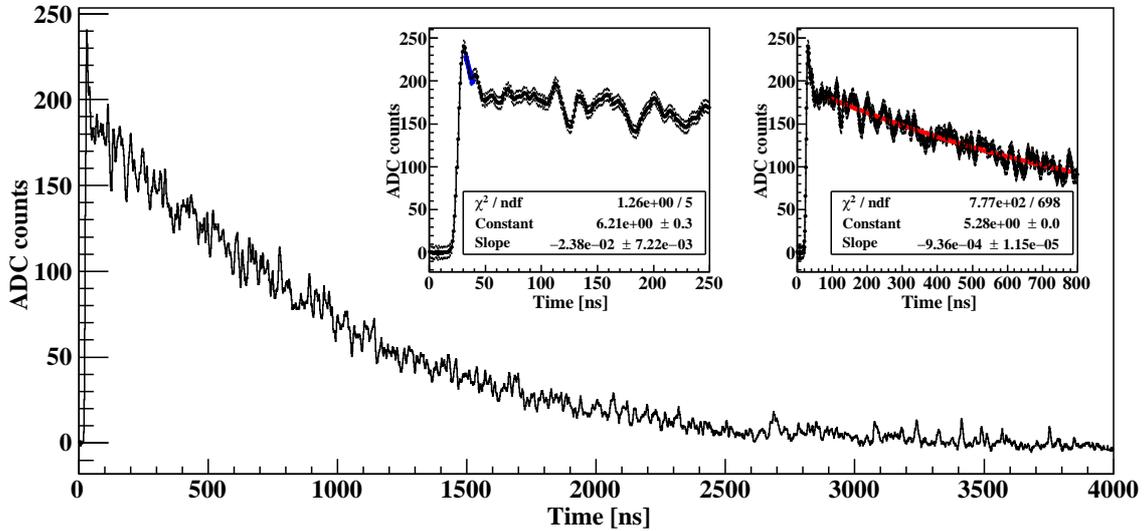}
  \caption{\label{f:lntpls}A typical light pulse taken at 77~Kelvin. The insets depict the fittings to the fast and slow decay components with eq.~\ref{e:exp}.}
\end{figure}

\begin{figure}[htpb]
  \includegraphics[width=0.49\linewidth]{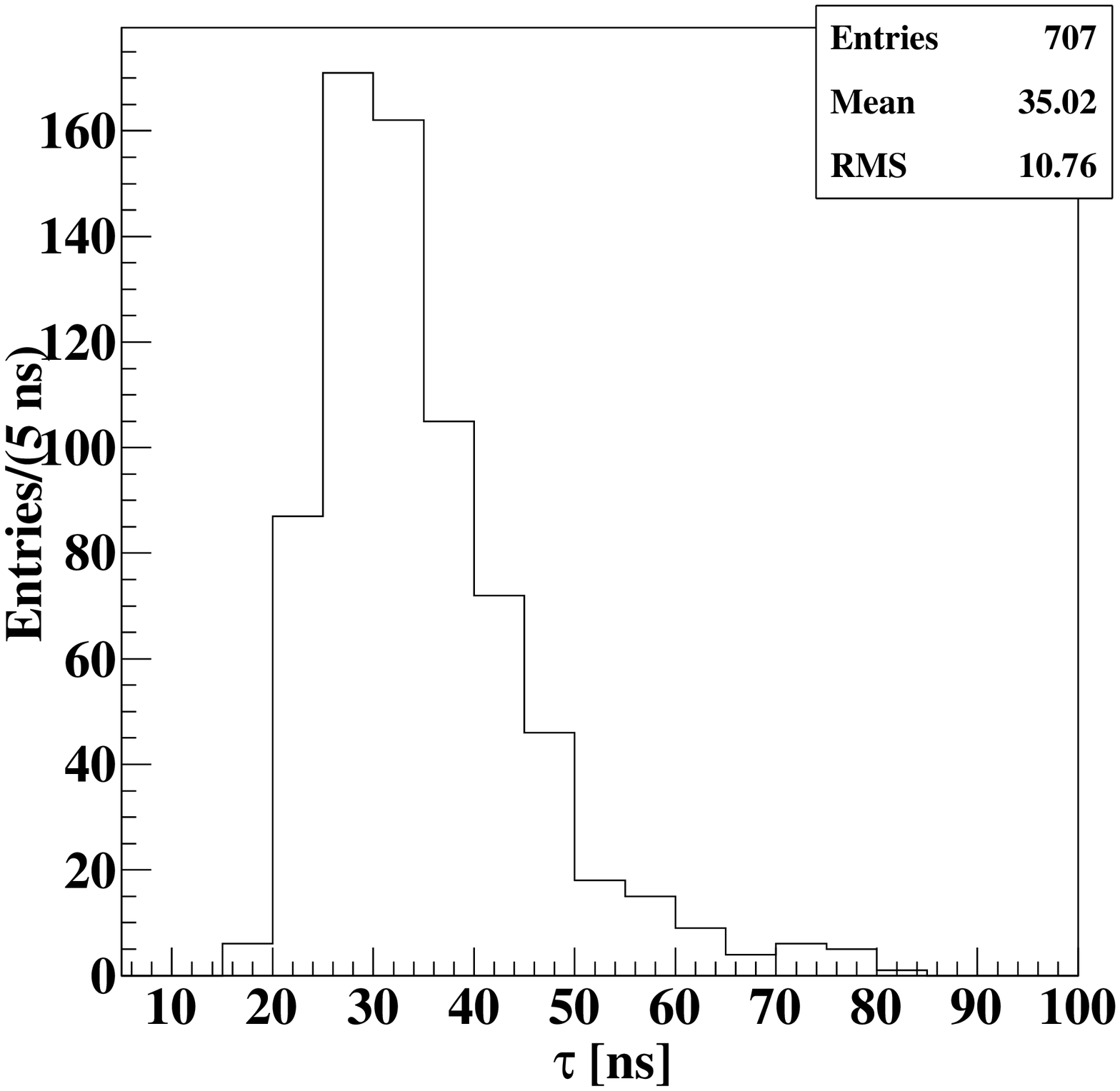}\hfill
  \includegraphics[width=0.49\linewidth]{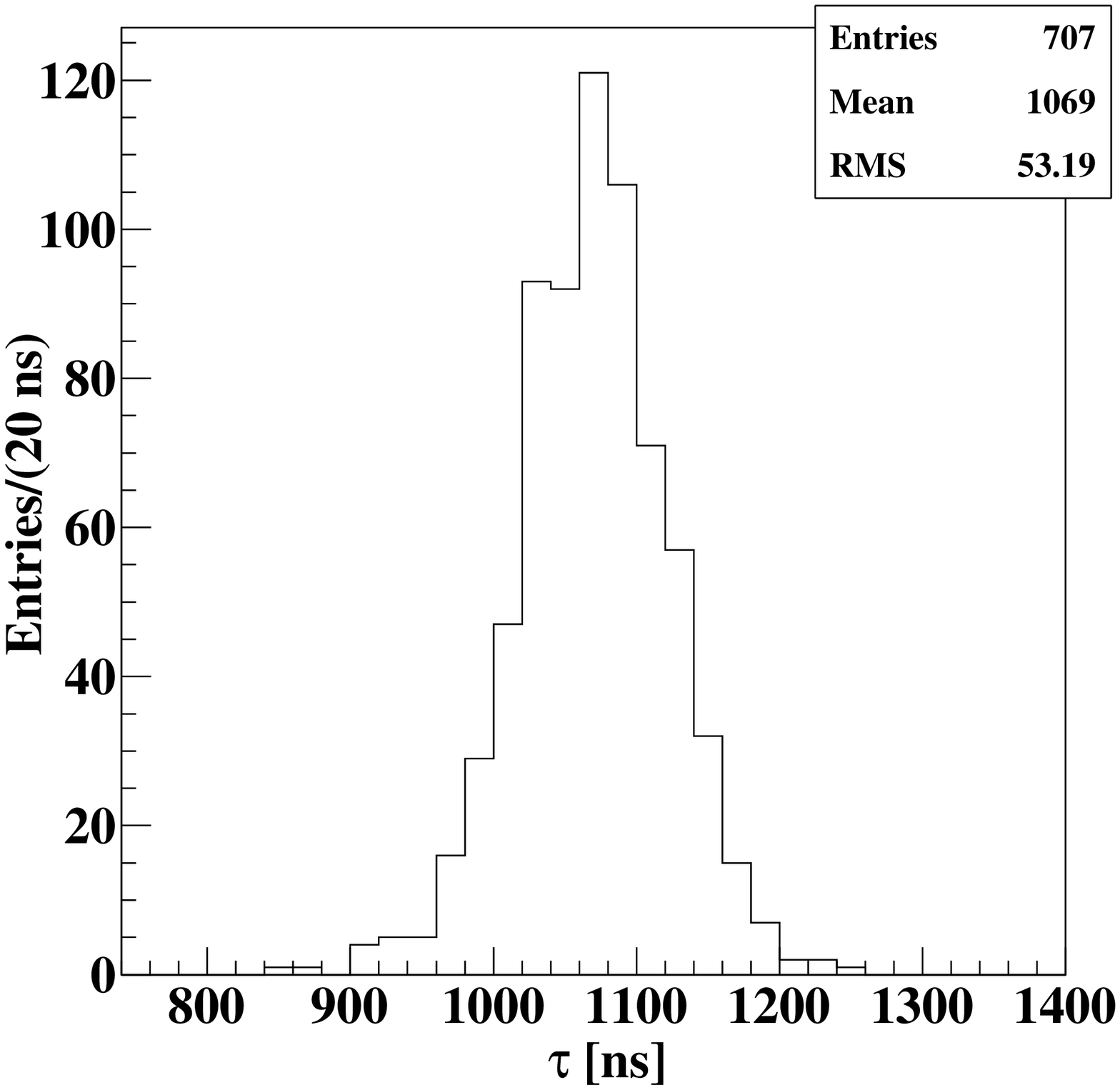}
  \caption{\label{f:lntdecay} Distributions of the fast(left) and slow(right) decay constants at 77~Kelvin.}
\end{figure}

A typical light pulse at 77~Kelvin is shown in figure~\ref{f:lntpls}. A fast and a slow decay component can be clearly identified and were fitted with eq.~\ref{e:exp}, as shown in the inlets of figure~\ref{f:lntpls}. The fitting window for the fast component is 7~ns, 3~ns after the maximum of the pulse, and the fitting window for the slow component is 700~ns, 55~ns after the maximum. The distributions of the two decay constants are shown in figure~\ref{f:lntdecay}. The mean of the fast/slow component distribution is 35/1069~ns with an RMS of 11/53~ns, respectively.

\begin{figure}[htpb]
  \includegraphics[width=\linewidth]{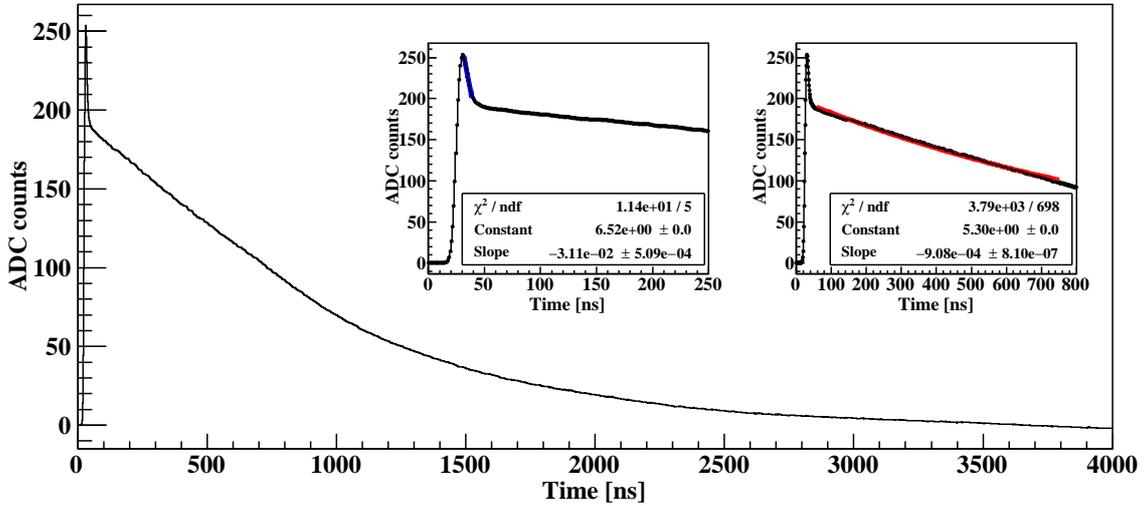}
  \caption{\label{f:lntsum}Averaged waveform of 800 events at 77~Kelvin. The inlets depict the exponential fittings to the fast and slow decay components with eq.~\ref{e:exp}.}
\end{figure}

The average of 707 individual pulses taken at 77~Kelvin was performed similarly to that at room temperature. The averaged waveform is shown in figure~\ref{f:lntsum}. The standard deviation of the ADC counts of the first 25 samples is 0.6, and was used as the uncertainty of each sample in the waveform. The same exponential fitting was performed and the fast and slow decay constants obtained this way were 32.2~ns and 1092~ns, consistent with those evaluated from individual pulses. The large $\chi^2$ of the fitting to the slow component indicates that some systematic uncertainties have not been taken into     account. Contamination from other decay components in the fitted time window may be one of them.

The decay constant of the slow component, $1069 \pm 53$~ns, measured at 77~Kelvin agrees perfectly with those mentioned in ref.~\cite{Bonanomi52, Nishimura95, Amsler02}. It is due to the 3.7~eV luminescence decay of self-trapped excitons in the crystal~\cite{Nishimura95}. The 2~ns fast component in ref.~\cite{Nishimura95} could not be observed because it was smeared out by the 15~ns width of single PE peaks as shown in figure~\ref{f:c1p3}, the 4.7~ns transit time spread of the PMT and the limited bandwidth of the amplifier used in the measurements.

The decay constants at root temperature reported in various literatures are not quite consistent~\cite{Bonanomi52, Kubota88, Nishimura95, Schotanus90, Amsler02}. They range from 2~ns to 1~$\mu$s. However, all of them agree that there are several components, most of which change with temperature. The inconsistency may be explained by slight temperature difference among the experiments, and the wavelength dependence of the detection efficiency of the light sensors used in their measurements. Ref.~\cite{Nishimura95} and eight references therein reported a decay constant of 15~ns. The CsI(pure) Data Sheet provided by the company, Saint-Gobain Crystals, also reports a primary decay constant of 16~ns~\cite{sg}. Our result is in perfect agreement with them.

\section{Impacts on rare-event search experiments}
The light yield achieved with this setup at 77~Kelvin, is about 4 times larger than that achieved in the KIMS-CsI experiment in 2014~\cite{kims14}, and is the largest in the world achieved with CsI to the authors' best knowledge. An energy threshold several times lower than that of KIMS-CsI can be reached assuming a similar background level. The sensitivity in the detection of dark matter particles using CsI can be significantly improved.

Since it is relatively easier to achieve a higher light collection efficiency using a smaller crystal, it makes more sense to compare the intrinsic light yield of the crystal, LY$_\text{crystal}$, instead of the light yield of the entire detection system, LY$_\text{system}$. LY$_\text{crystal}$ can be calculated using eq~\ref{e:ly}. Unfortunately, the light collection efficiency $\epsilon$ in the equation could not be measured with the current setup. Assuming $\epsilon=100\%$, LY$_\text{crystal}$ can be estimated as $(20.4~\text{PE/keV})/27.3\% \approx 75~\text{photons/keV}$, which is higher than that of CsI(Tl) at room temperature. Assuming $\epsilon=90\%$, LY$_\text{crystal}$ becomes $$(20.4~\text{PE/keV})/90\%/27.3\%\approx 83~\text{photons/keV}.$$ This is about 2 times more than those of NaI(Tl), CsI(Tl) and CsI(Na) at room temperature, but is in agreement with the results given in ref.~\cite{Bonanomi52, Sciver56, Moszynski03, Moszynski05}. Undoped CsI presents most of the advantages listed in ref.~\cite{cosi15} for CEvNS detection. In addition, it has a higher light yield than CsI(Na) studied in ref.~\cite{cosi15}, indicating a possibly lower energy threshold. Undoped CsI at 77~Kelvin presents a fast (2~ns) and a slow (1~$\mu$s) decay components, suggesting possible discrimination of nuclear and electronic recoil events using pulse shape analysis. No decay constant measured with neutron sources was found in the literature. However, a measurement with $\alpha$-particles did show a different decay constant of 500~ns~\cite{Hahn53}, further suggesting the possibility.

Liquid nitrogen or argon can be used to cool an array of undoped CsI crystals directly coupled with PMTs and to shield them from external radiation at the same time. They are relatively cheap and are easy to purify. Liquid argon is also a scintillation material and can be used as an anti-coincident veto system. Such a setup is a possible alternative technique for both dark matter and CEvNS detection, providing mechanisms to lower the background level and energy threshold. Given similar behavior of undoped NaI as given in the literature~\cite{Sciver58, Sciver60, Moszynski03}, both NaI and CsI can be deployed in such a setup, providing a comprehensive verification of the DAMA result.

It has been suggested in a recent arXiv article~\cite{derenzo16} that pure NaI and CsI at cryogenic temperature can be used to detect sub-GeV dark matter particles interacting with electrons instead of nuclei.The energy threshold can be very low if the system is allowed to be triggered by one or two scintillation photons. The major background is the dark noise of the photon sensors deployed. PMTs mentioned in our work typically have a dark count rate of several tens Hz at 77~Kelvin. If a coincident trigger of two PMTs watching the same crystal is required, the background from the dark count is negligible. This points out a much broader application of our technique than the verification of the DAMA result.

\section{Conclusion}
A light yield of $20.4 \pm 0.8$~PE/keV was achieved with an undoped CsI crystal directly coupled with a PMT that works at 77~Kelvin. This is by far the largest in the world achieved with CsI crystals to the authors' best knowledge. The scintillation decay constants of the crystal were measured at both room temperature and 77~Kelvin. They agree with those in the literature, confirming the crystal used in the study was indeed an undoped CsI. Possible experimental setup has been suggested to use this technique for both dark matter and CEvNS detection.

\acknowledgments
The authors thank Christina Keller for her careful reading of manuscript. The laboratory spaces and part of equipment used for conducting this study was kindly provided by the XMASS collaboration. This work was supported by NSF PHY-1506036, the Office of Research at the University of South Dakota, the Grant-in-Aid for Encouragement of Young Scientists (B) project No. 26800122, MEXT, Japan and the World Premier International Research Center Initiative (WPI Initiative), MEXT, Japan. Computations supporting this project were performed on the High Performance Computing systems at the University of South Dakota. We thank its manager, Doug Jennewein, for providing valuable technical expertise to this project.

\bibliography{ref}
\bibliographystyle{JHEP}
\end{document}